\documentclass[runningheads,a4paper]{llncs}

\usepackage{cmap}
\usepackage[T1]{fontenc}
\usepackage{multirow}
\usepackage{graphicx}
\usepackage[normalem]{ulem}
\usepackage[ngerman,english]{babel}
\addto\extrasenglish{\languageshorthands{ngerman}\useshorthands{"}}

\usepackage[%
rm={oldstyle=false,proportional=true},%
sf={oldstyle=false,proportional=true},%
tt={oldstyle=false,proportional=true,variable=true},%
qt=false%
]{cfr-lm}
\usepackage[math]{blindtext}

\usepackage{cite}

\usepackage{paralist}

\usepackage{csquotes}

\usepackage{microtype}
\usepackage{arabtex}
\usepackage{utf8}
\setcode{utf8}

\usepackage{url}
\makeatletter
\g@addto@macro{\UrlBreaks}{\UrlOrds}
\makeatother

\usepackage{xcolor}

\usepackage{pdfcomment}
\hypersetup{hidelinks,
   colorlinks=true,
   allcolors=black,
   pdfstartview=Fit,
   breaklinks=true}
\usepackage[all]{hypcap}

\usepackage[capitalise,nameinlink]{cleveref}
\crefname{section}{Sect.}{Sect.}
\Crefname{section}{Section}{Sections}

\usepackage{xspace}

\DeclareFontFamily{U}{MnSymbolC}{}
\DeclareSymbolFont{MnSyC}{U}{MnSymbolC}{m}{n}
\DeclareFontShape{U}{MnSymbolC}{m}{n}{
    <-6>  MnSymbolC5
   <6-7>  MnSymbolC6
   <7-8>  MnSymbolC7
   <8-9>  MnSymbolC8
   <9-10> MnSymbolC9
  <10-12> MnSymbolC10
  <12->   MnSymbolC12%
}{}
\DeclareMathSymbol{\powerset}{\mathord}{MnSyC}{180}

\begin{document}

\title{Trump vs. Hillary: What went Viral during the 2016 US Presidential Election}
\titlerunning{Trump vs. Hillary}  
%
\author{Kareem Darwish\inst{1} \and Walid Magdy\inst{2} \and Tahar Zanouda\inst{1}}
\authorrunning{Kareem Darwish et al.} 
%
\tocauthor{Kareem Darwish, Walid Magdy, Tahar Zanouda}
\institute{Qatar Computing Research Institute, HBKU, Doha, Qatar\\
\email{\{kdarwish,tzanouda\}@hbku.edu.qa},\\ 
\and School of Informatics, The University of Edinburgh, Scotland\\
\email{wmagdy@inf.ed.ac.uk}
}

\maketitle              
\vspace{-0.8cm}
\begin{abstract}
In this paper, we present quantitative and qualitative analysis of the top retweeted tweets (viral tweets) pertaining to the US presidential elections from September 1, 2016 to Election Day on November 8, 2016. For everyday, we tagged the top 50 most retweeted tweets as supporting or attacking either candidate or as neutral/irrelevant. Then we analyzed the tweets in each class for: general trends and statistics; the most frequently used hashtags, terms, and locations; the most retweeted accounts and tweets; and the most shared news and links. In all we analyzed the 3,450 most viral tweets that grabbed the most attention during the US election and were retweeted in total 26.3 million times accounting over 40\% of the total tweet volume pertaining to the US election in the aforementioned period. Our analysis of the tweets highlights some of the differences between the social media strategies of both candidates, the penetration of their messages, and the potential effect of attacks on both.
\end{abstract}

\vspace{-0.8cm}
\section{Introduction}
\vspace{-0.3cm}
Social media is an important platform for political discourse and political campaigns \cite{shirky2011political,west2013air}. Political candidates have been increasingly using social media platforms to promote themselves and their policies and to attack their opponents and their policies. Consequently, some political campaigns have their own social media advisers and strategists, whose success can be pivotal to the success of the campaign as a whole. The 2016 US presidential election is no exception, in which the Republican candidate Donald Trump won over his main rival Hilary Clinton. 

In this work, we showcase some of the Twitter trends pertaining to the US presidential election by analyzing the most retweeted tweets in the sixty eight days leading to the election and election day itself. In particular, we try to answer the following research questions:

1) Which candidate was more popular in the viral tweets on Twitter? What proportion of viral tweets in terms of number and volume were supporting or attacking either candidate?

2) Which election related events, topics, and issues pertaining to each candidate elicited the most user reaction, and what were their effect on each candidate?
Which accounts were the most influential?

3) How credible were the links and news that were shared in viral tweets?

To answer these questions, we analyze the most retweeted (viral) 50 tweets per day starting from September 1, 2016 and ending on Election Day on November 8, 2016. The total number of unique tweets that we analyze is 3,450, whose retweet volume of 26.3 million retweets accounts for over 40\% of the total tweets/retweets volume concerning the US election during that period. We have manually tagged all the tweets in our collection as supporting or attacking either candidate, or neutral (or irrelevant). For the different classes, we analyze retweet volume trends, top hashtags, most frequently used words, most retweeted accounts and tweets, and most shared news and links.
Our analysis of the tweets shows some clear differences between the social media strategies of each candidate and subsequent user responses. For example, our observations show that the Trump campaign seems to be more effective than the Clinton campaign in achieving better penetration and reach for: the campaign's slogans, attacks against his rival, and promotion of campaign activities in different US states. We also noticed that the prominent vulnerabilities for each candidate were Trump's views towards women and Clinton's email leaks and scandal. In addition, our analysis shows that the majority of tweets benefiting Clinton were actually more about attacking Trump rather than supporting Clinton, while for Trump, tweets in his favor had more balance between supporting him and attacking Clinton. By analyzing the links in the viral tweets, the tweets attacking Clinton had the most number of links (accounting for 58\% of the volume of shared links), where approximately half were to highly credible sites and the remaining were to sites of mixed credibility.
We hope that our observations and analysis would aid political and social scientists in understanding some of the factors that may have led to the eventual outcome of the election.
\section{Background}
\label{sec:related}
\vspace{-0.2cm}%
Social media is a fertile ground for developing tools and algorithms that can capture the opinions of the general population at a large scale \cite{jungherr2015analyzing}. Much work has studied the potential possibility of predicting the outcome of political elections from social data. Yet, there is no unified approach for tackling this problem. While Bollen et al. \cite{bollen2011modeling} analyzed people's emotions (not sentiment) towards the US 2008 Presidential campaign, the authors used both US 2008 Presidential campaign and election of Obama as a case study. Sentiment typically signifies preconceived positions towards an issue, while emotions, such as happiness or sadness, are temporary responses to external stimulus. Though the authors mentioned the feasibility of using such data to predict election results, they did not offer supporting results. Using the same approach, O'Connor et al.\cite{o2010tweets} discussed the feasibility of using Twitter data to replace polls. Tumasjan et al. \cite{tumasjan2010predicting} provided one of the earliest attempts for using this kind of data to estimate election results. They have used twitter data to forecast the national German federal election, and investigated whether online interactions on Twitter validly mirrored offline political sentiment. The study was criticized for being contingent on arbitrary experimental variables  \cite{jungherr2012pirate,metaxas2011not,gayo2011limits,gayo2011don}. Metaxas et al. \cite{metaxas2011not} argued that the predictive power of Twitter  is exaggerated and it cannot be accurate unless we are able to identify unbiased representative sample of voters. However, these early papers kicked off a new wave of research initiatives that focus on studying the political discourse on Twitter, and how this social platform can be used as a proxy for understanding people's opinions \cite{mislove2011understanding}.
On the other hand, many studies have questioned the accuracy, and not just the feasibility, of using social data to predict and forecast \cite{gayo2011limits,gayo2011don}. Without combining the contextual information, together with social interactions, the results might lead to biased findings. One of the main problems of studying political social phenomena on Twitter is that users tend to interact with like-minded people, forming so-called ``echo chambers'' \cite{barbera2015tweeting,colleoni2014echo,magdy2016isisisnotislam}. Thus, the social structure of their interactions will limit their ability to interact in a genuinely open political space. Though the literature on the predictive power of twitter is undoubtedly relevant, the focus of this work is on performing post-election analysis to better understand the potential strengths and weaknesses of the social media strategies of political campaigns and how they might have contributed to eventual electoral outcomes.

A step closer to our politically motivated work, we find some studies that focused on studying the US presidential election. For the 2012 US presidential election, Shi et al. \cite{shi2012predicting}  analyzed millions of tweets to predict the public opinion towards the Republican presidential primaries. The authors trained a linear regression model and showed good results compared to pre-electoral polls during the early stages of the campaign. In addition, the Obama campaign data analytics system\footnote{\url{edition.cnn.com/2012/11/07/tech/web/obama-campaign-tech-team/index.html}}, a decision-making tool that harnesses different kind of data ranging from social media to news, was developed to help the Obama team sense the pulse of public opinion and strategically manage the campaign. With an attempt to study the recent US presidential election, early studies have showed promising results. For example, Wang et al. \cite{wang2016deciphering} studied the growth pattern of Donald Trump's followers at a very early stage. They characterized individuals who ceased to support Hillary Clinton and Donald Trump, by focusing on three dimensions of social demographics: social status, gender, and age. They also analyzed Twitter profile images to reveal the demographics of both Donald Trump and Hillary followers. Wang et al. \cite{wang2016catching} studied the frequency of `likes' for every tweet that Trump published. More recently, Bovet et al. \cite{bovet2016predicting} developed an analytical tool to study the opinion of Twitter users, in regards to their structural and social interactions, to infer their political affiliation. Authors showed that the resulting Twitter trends follow the New York Times National Polling Average.

Concerning disseminating information, recent research has focused on how candidates engaged in spreading fake news and in amplifying their message via the use of social bots, which are programmatically controlled accounts that produce content and automatically interact with other users \cite{kollanyi2016bots, davis2016botornot}. Bessi and Ferrara \cite{bessi2016social} investigated how the presence of social media bots affected political discussion around the 2016 U.S. presidential election. Authors suggest that the presence of social media bots can negatively affect democratic political discussion rather than improving it, which in turn can potentially alter public opinion and endanger the integrity of the presidential election. In the same context, Giglietto et al. \cite{giglietto2016fakes} studied the role played by ``fake-news'' circulating on social media during the 2016 US Presidential election.
In this study, we provide a quantitative and qualitative analysis of tweets related to the 2016 US presidential election. By tapping into the wealth of Twitter data, we focus on analyzing and understanding the possible factors underlying the success and failure of candidates. Our work builds upon the aforementioned research in social and computer sciences to study and measure the volume and diversity of support for the two main candidates for the 2016 US presidential elections, Donald Trump and Hillary Clinton.

\vspace{-0.3cm}%
\section{Data Collection and Labeling}
\vspace{-0.3cm}%
To acquire the tweets that are relevant to the US presidential election, we obtained the tweets that were collected by \textit{TweetElect.com} 
from September 1, 2016 to November 8, 2016 (Election Day). TweetElect is a public website that aggregates tweets that are relevant to the 2016 US presidential election. It shows the most retweeted content on Twitter including text tweets, images, videos, and links. The site uses state-of-the-art adaptive filtering methods for detecting relevant tweets on broad and dynamic topics, such as politics and elections \cite{magdy2014adaptive,magdy2016unsupervised}. TweetElect used an initial set of 38 keywords related to the US elections for streaming relevant tweets. Consequently, adaptive filtering continuously enriches the set of keywords with additional terms that emerge over time \cite{magdy2016unsupervised}. The 38 seeding keywords included all candidate names and common keywords (including hashtags) about the elections and participating parties.
During the period of interest, the total number of aggregated tweets per day (including retweets) related to the US elections typically ranged between 200K and 600K. This number increased dramatically after specific events or revelations, such as after the presidential debates and Election Day, when the number of tweets exceeded 3.5 million tweets. The total number of unique tweets collected by TweetElect between September 1 and November 8 was 6.8 million, while the full volume of tweets including retweets was 65.8 million.

\begin{figure}[t]
\centering
\includegraphics [width=\linewidth]{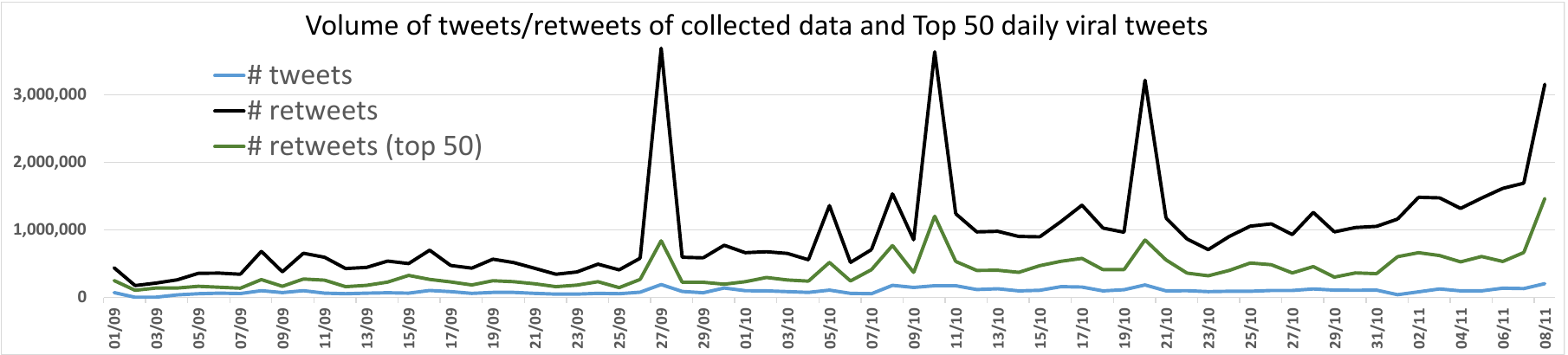}
\vspace{-0.5cm}
\caption{Volume of retweets the top 50 daily viral tweets on the US collection compared to the full volume of tweets and retweets.}
\label{vol-of-retweets}
\end{figure}

In this work, we are interested in analyzing the most ``viral'' tweets pertain to the US presidential elections, as they typically express the topics that garnered the most attention on Twitter~\cite{magdy2016trump}. Specifically, we constructed a set of the most retweeted 50 tweets for everyday in the period of interest. Thus, our collection contained 3,450 unique tweets that were retweeted 26.6 million times, representing more than 40\% of the total volume of tweets during that period.
Out of the 3,450 tweets in our collection, 700 were authored by the official account of Donald Trump, accounting for 8.7 million retweets, and 698 were authored by Clinton's official account, accounting for 4.7 million retweets.
Figure \ref{vol-of-retweets} shows the distribution of the number of tweets collected by TweetElect in the US elections in the period of study. The number of unique tweets, retweets volume, and retweets volume of the top viral daily tweets are displayed. As shown, the volume of tweets increased as election day approached. The biggest four peaks in the graph represent the days following the presidential debates and the election day. As it is shown, the retweet volume of the top 50 daily viral tweets correlates well with the full retweet volume, with a Pearson correlation of 0.92, which indicates nearly identical trend.

We calculated the daily coverage $C_{daily}$ of the top 50 daily viral tweets to the full tweet volume. The daily coverage ranged between 23\% and 66\%, with the majority of the days having $C_{daily}$ over 40\%. This indicates that the top 50 viral daily tweets may offer good coverage and reasonable indicators for public interaction with the US election on Twitter.

All tweets were labeled with one of five class labels, namely: ``support Trump'', ``attack Trump'', ``support Clinton'', ``attack Clinton'', or ``neutral/irrelevant'', with tweets being allowed to have multiple labels if applicable. Support for a candidate included praising or defending the candidate, his/her supporters, or staff, spreading positive news about the candidate, asking people to vote for the candidate, mentioning favorable polls where the candidate is ahead, promoting the candidate's agenda, or advertising appearances such as TV interviews or rallies. Attacking a candidate included maligning and name calling targeted at the candidate, his/her supporters, or staff, spreading negative news about the candidate, mentioning polls where the candidates is behind, or attacking the candidate's agenda. Other tweets were labeled as neutral/irrelevant. Tweets were allowed to have more than one label such as ``support Trump, attack Clinton''. The labeling was done by an annotator with strong knowledge of US politics. The annotator was instructed to check the content of tweets carefully including any images, videos, or external links to obtain accurate annotations. In addition, we advised the annotator to check the profile of tweet authors to better understand their position towards the candidates if needed.  One of the authors took a random sample of 50 tweets to verify the correctness of the annotation.  In all, both agreed fully on 90\% of the sample, partially agreed on 8\%, and disagreed on the remaining 2\%.  The agreement between the annotator and the author, as measured using Cohen's Kappa, is 0.87, meaning nearly perfect agreement.  Table \ref{table:exampleAnnotations} shows few sample labeled tweets. 

\begin{table}[t]
\tiny
\begin{center}
\begin{tabular}{p{.18\linewidth}|p{.82\linewidth}}
Label & Tweet \\ \hline
attack Trump, support Clinton & Donald Trump is unfit for the office of president. Fortunately, there's an exceptionally qualified candidate @HillaryClinton \\ \hline
attack Trump, attack Clinton & Donald Trump looks like what Hillary Clinton smells like\\ \hline
attack Clinton & A rough night for Hillary Clinton ABC News. \\ \hline
attack Trump & Trump to Matt Lauer on Iraq: I was totally against the war. Here's proof Trump is lying: \url{https://t.co/6ZhgJMUhs3} \\ \hline
neutral & It s official: the US has joined the \#ParisAgreement \url{https://t.co/qYN1iRzSJk} \\ \hline
\end{tabular}
\end{center}
\vspace{-0.3cm}
\caption{Example annotations}
\label{table:exampleAnnotations}
\end{table}

\section{Tweet Analysis}

\begin{table}[t]
\tiny
\begin{center}
\setlength\tabcolsep{2.5pt}
\begin{tabular}{l|r|r|r|r}
Class & No. of Tweets & \% of Tweets & No. of Retweets & \% of Retweets\\ \hline
Support Clinton 	&	506	&	13.6\%	&	3,205,303	&	11.5\%	\\	
Attack Trump 	&	712	&	19.1\%	&	6,373,549	&	22.8\%	\\	
Support Trump 	&	848	&	22.8\%	&	6,896,940	&	24.7\%	\\	
Attack Clinton 	&	1,458	&	39.2\%	&	9,441,921	&	33.8\%	\\	
Neutral/irrelevant 	&	199	&	5.3\%	&	1,978,784	&	7.1\%	\\	\hline
\end{tabular}
\caption{Number of tweets and retweet volume per class.}
\label{table:retweetVolumePerClass}
\end{center}
\end{table}

\begin{figure}[t]
\begin{center}
\vspace{-1cm}
\includegraphics[width=\textwidth]{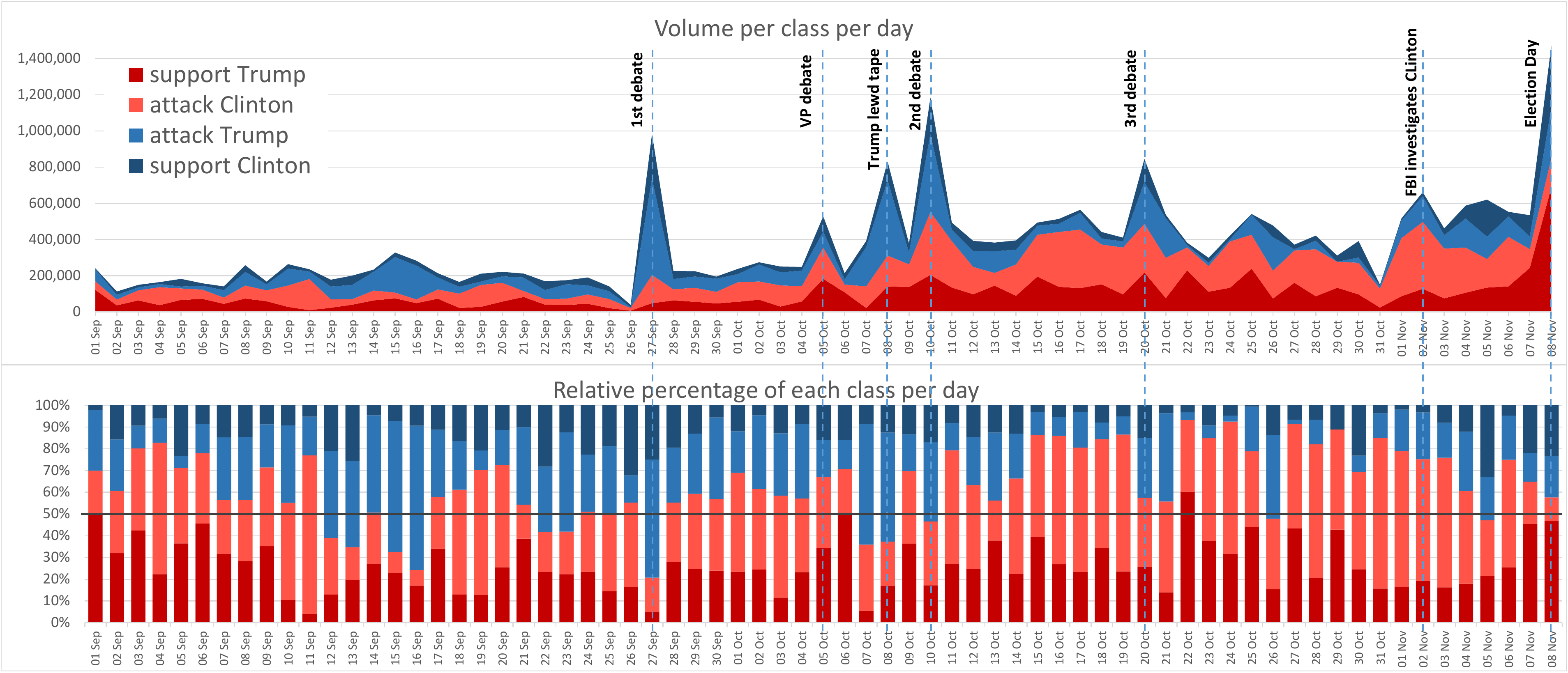}
\vspace{-0.3cm}
\caption{(a) Retweet volume for each label per day, and (b) Percentage of labels per day.}
\label{figure:perDayPlots}
\end{center}
\end{figure}

\vspace{-0.2cm}%
We analyze tweets in every class from different perspectives. Specifically, we look at: user engagement with class over time, most viral events during election, most discussed topics, most shared links and news, and most influential accounts supporting them. \\
\textbf{\textit{Popularity of Candidates on Twitter}} Table \ref{table:retweetVolumePerClass} shows the number of tweets and their retweet volume for each label. For further analysis, we ignore tweets that are labeled as neutral/irrelevant as they are not interesting to this work. Figure~\ref{figure:perDayPlots} shows a break down for each label across all days.
Table \ref{table:retweetVolumePerClass} shows that the majority of the tweets and retweets volume were in favor of Trump (61.9\% of tweets, 58.6\% of retweet volume), either by supporting him or attacking Clinton, while those in favor of  Clinton represent only (32.7\% of tweets, 34.3\% of retweet volume) of the total, either supporting her or attacking Trump. 
It can be observed that the retweet volume of tweets attacking Clinton outnumbered the retweet volume of tweets supporting her by nearly a 3-to-1 margin, while for Trump those supporting and attacking him were almost evenly matched.
Figure \ref{figure:perDayPlots}(a) shows the distribution of support/attack of each candidate over the period of study. Figure \ref{figure:perDayPlots}(b) shows the relative per day retweet volume for tweets supporting/attacking each candidate. As expected, large spikes in volume happen in conjunction with major events such the presidential and vice presidential debates, Election Day, the release of Trump's lewd Access Hollywood tape, and the FBI announcement concerning the reopening of the investigation of Clinton. An interesting observation from Figure \ref{figure:perDayPlots}(b) is that tweets in favor of Trump (either supporting him or attacking Clinton) were retweeted more than tweets in favor of Clinton for 85\% of the days, with the exception of a few days, especially the day following the first presidential debate and following the release of Trump's lewd tape. This observation might be different to the trends in normal media, where large number of articles were more negative towards Trump \cite{van2017leading}. 


\begin{table}[ht!]
\tiny
\begin{center}
\begin{tabular}{r|l|r}
\hline
\multicolumn{3}{c}{support Clinton} \\ \hline
Date & Leading topic & \% \\ \hline
11/5 & release of a video promoting Clinton &	32.9\% \\
9/26 & after first presidential debate & 32.2\%	\\
9/22 & interview w/ comedy program (two Ferns) & 28.2\%	\\
 \hline
\multicolumn{3}{c}{attack Clinton} \\ \hline
9/11 & Clinton faints at 9/11 memorial	& 72.9\%\\
10/31 & \begin{minipage}[t]{.7\columnwidth} WikiLeaks: Clinton got gifts from foreign governments \end{minipage} & 69.4\% \\
10/19 & WikiLeaks: Clinton got gift from Qatar & 62.9\% \\ \hline
\multicolumn{3}{c}{support Trump} \\ \hline
10/22 & Trump leads in 3 national polls	& 60.2\%\\
10/6 & Trump leads in Virginia poll	& 50.7\% \\
9/1 & Trump says to build wall with Mexico & 50.1\% \\ \hline
\multicolumn{3}{c}{attack Trump} \\ \hline
9/16 & Sanders attacks Trump on immigration	& 66.4\%\\
9/15 & \begin{minipage}[t]{.7\columnwidth} Trump quoted: ``pregnancy very inconvenient for business'' \end{minipage}	& 60.2\% \\
10/7 & Trump lewd tape released & 55.5\% \\
 \hline
\end{tabular}
\caption{Top 3 days when relative volume for each class peaked with leading topic and percentage.}.
\label{table:topDaysForClasses}
\vspace{-0.3cm}
\end{center}
\end{table}
\begin{table}[ht]
\tiny
\begin{center}
\setlength\tabcolsep{2.5pt}
\begin{tabular}{l|r|l}
\hline
\multicolumn{3}{c}{\textbf{Support Clinton}} \\ \hline
Category & \#tag Freq.  & Hashtags \\ \hline
Debate performance &	486,402 & \begin{minipage}[t]{.65\columnwidth} \#DebateNight	\#Debate	\#SheWon	\#VPDebate	\#Debates2016	\#NBCNewsForum \end{minipage}	 \\
Attacking Trump & 114,614 & \begin{minipage}[t]{.65\columnwidth} \#TrumpTapes	\#ImVotingBecause	\#TangerineNightmare	\#InterrogateTrump	\#ImWithTacos	\#AlSmithDinner \end{minipage} \\
Get out the vote & 86,297 & \begin{minipage}[t]{.65\columnwidth} \#ElectionDay	\#Election2016	\#Voting	\#OHVotesEarly	\#Vote	\#Elections2016 \end{minipage} \\
Campaign Issues & 15,055 & \begin{minipage}[t]{.65\columnwidth} \#NationalComingOutDay	\#LatinaEqualPay	\end{minipage} \\
Campaign slogan & 10,084 & \begin{minipage}[t]{.65\columnwidth} \#ImWithHer	\end{minipage} \\ \hline \hline
\multicolumn{3}{c}{\textbf{Attack Clinton}} \\ \hline
Corruption/lying &	842,892 &	\begin{minipage}[t]{.65\columnwidth} \#DrainTheSwamp	\#CrookedHillary	\#BigLeagueTruth	\#GoldmanSach	\#FollowTheMoney	\#PayToPlay	\end{minipage}\\
Wikileaks Releases	& 640,330 &	\begin{minipage}[t]{.65\columnwidth} \#PodestaEmails	\#Wikileaks	\#PodestaEmails\{8,31,28,26,15\}	\#SpiritCooking	\#Podesta	
\#DNCLeak	\#DNCLeak2	
\#FreeJulian	
\#AnthonyWeiner \end{minipage} \\
Debate performance	& 348,749 & \begin{minipage}[t]{.65\columnwidth}	\#Debates2016	\#Debate	\#DebateNight	\#VPDebate	\#Debates \end{minipage}\\
Health Care	& 113,318 &	\#ObamaCare	\#ObamaCareFail	\#ObamaCareInThreeWords	\\
\begin{minipage}[t]{.2\columnwidth} Tension w/Sanders supporters \end{minipage}	 & 86,333 &	\#FeelTheBern	\#BasementDwellers \\
Media/Election Bias	 & 78,511 &	\#VoterFraud	\#RiggedSystem			\\								
Ben Ghazi attack	& 46,725 &	\#BenGhazi\\
Clinton's health	 & 30,747 &	\#HillarysHealth\\
\hline \hline
\multicolumn{3}{c}{\textbf{Support Trump}} \\ \hline
Campaign slogans	 & 2,089,162 &	\begin{minipage}[t]{.65\columnwidth} \#MAGA (Make America Great Again)	\#DrainTheSwamp	\#AmericaFirst	\#MakeAmericaGreatAgain	\#ImWithYou	\#TrumpTrain	\#TrumpPence16	\end{minipage} \\
Get out the vote	& 349,078 & \begin{minipage}[t]{.65\columnwidth}	\#VoteTrumpPence16	\#ElectionDay	\#VoteTrump	\#Vote	\#IVoted	\#ElectionNight	\#TheHoodForTrump	\#Vote2016	\#EarlyVote \end{minipage}\\
Campaign news	 & 183,031 & \begin{minipage}[t]{.65\columnwidth}	\#ICYMI (in case you missed it)	\#TrumpRally	\#Gettysburg (public speech) \end{minipage}	\\
Debate performance	& 74,808 &	\#VPDebate	\#DebateNight	\#Debates	\#Debates2016					\\
Attacking Clinton	& 66,318 & \begin{minipage}[t]{.65\columnwidth} \#FollowTheMoney	\#DNCLeak	\#ObamaCareFailed	\#BigLeagueTruth \end{minipage} \\ \hline \hline	
\multicolumn{3}{c}{\textbf{Attack Trump}} \\ \hline
Debate performance	& 776,956 &	\#DebateNight	\#Debate	\#VPDebate	\#Debates	\#NBCNewsForum	\\
Sexual misconduct	& 78,043 &	\#TrumpTapes	\#NatashaStoynoff \\
\begin{minipage}[t]{.2\columnwidth} General attacks and insults	\end{minipage} & 44,100 & \begin{minipage}[t]{.65\columnwidth}	\#ImVotingBecause	\#TangerineNightmare 	\#Lunatic	\#DangerousMan	\#Unfit	\#Deceit	\#InterrogateTrump \end{minipage} \\
Comedy attacks	& 26,980 &	\begin{minipage}[t]{.65\columnwidth} \#ACloserLook (segment on Seth Meyer comedy show)	\#AlSmithDinner (charity dinner)	\end{minipage} \\
\begin{minipage}[t]{.2\columnwidth} Attacking Trump's speech \end{minipage}	& 21,117 &	\#LoveTrumpsHate	\#NastyWoman \\ \hline
\end{tabular}
\caption{Top hashtags attacking or supporting both candidates with their categories.}
\label{table:topHashTagsWithCat}
\vspace{-0.3cm}
\end{center}
\end{table}

\begin{figure}[bt]
\centering
\includegraphics[width=.6\linewidth]{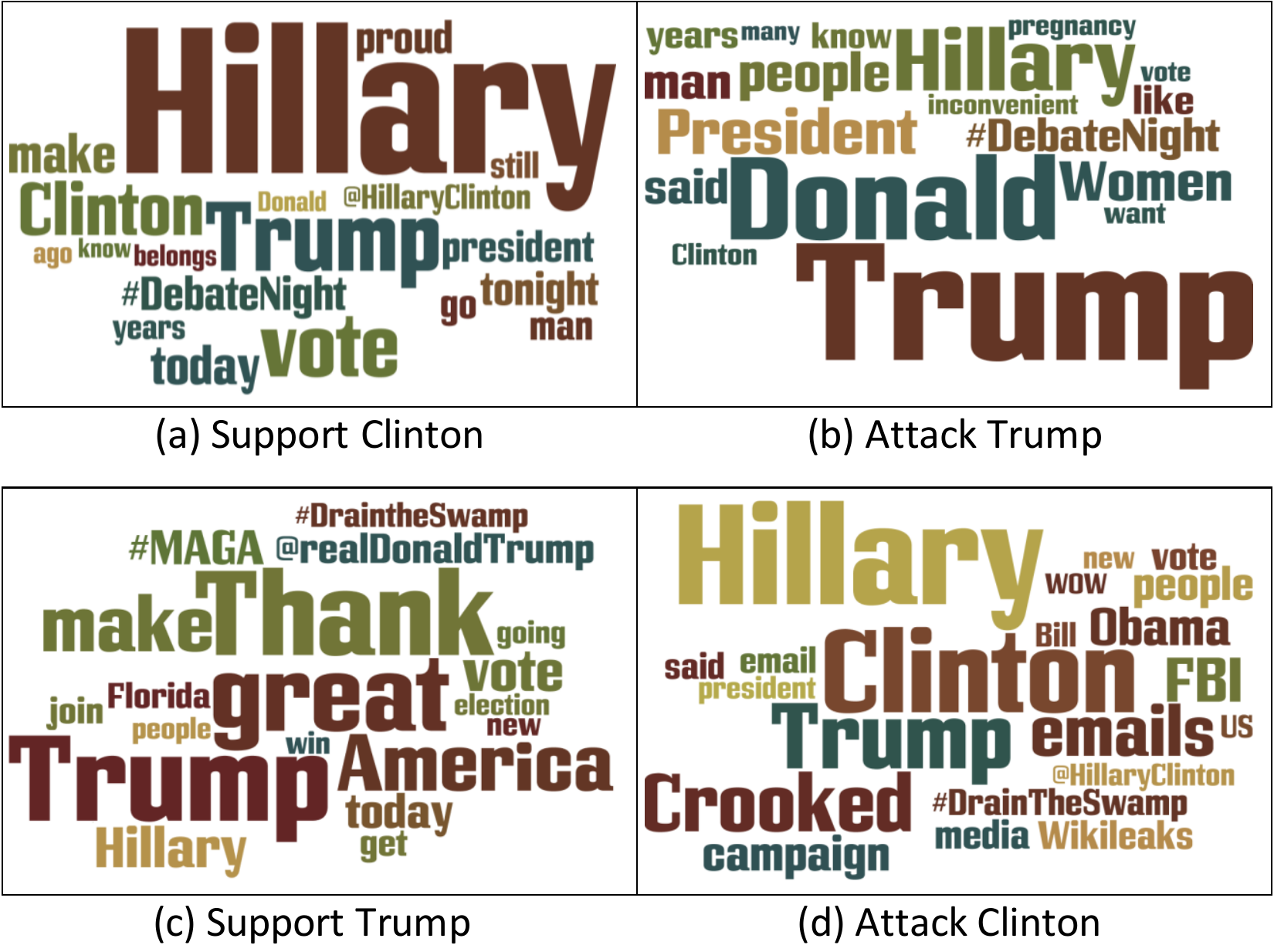}
\vspace{-0.3cm}
\caption{Tag-cloud of top frequent terms in each class support/attack Clinton/Trump.}
\label{tag-cloud}
\vspace{0.5cm}
\end{figure}

\begin{figure}
    \centering
    \begin{minipage}{0.5\textwidth}
        \centering
\includegraphics[width=\linewidth]{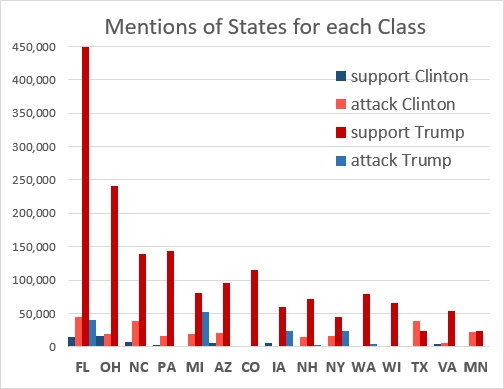}
\label{state-mentions}
\vspace{-0.5cm}
\caption{Mentions of states for each class.}
    \end{minipage}\hfill
    \begin{minipage}{0.5\textwidth}
            \vspace{8pt}
\includegraphics[width=\linewidth]{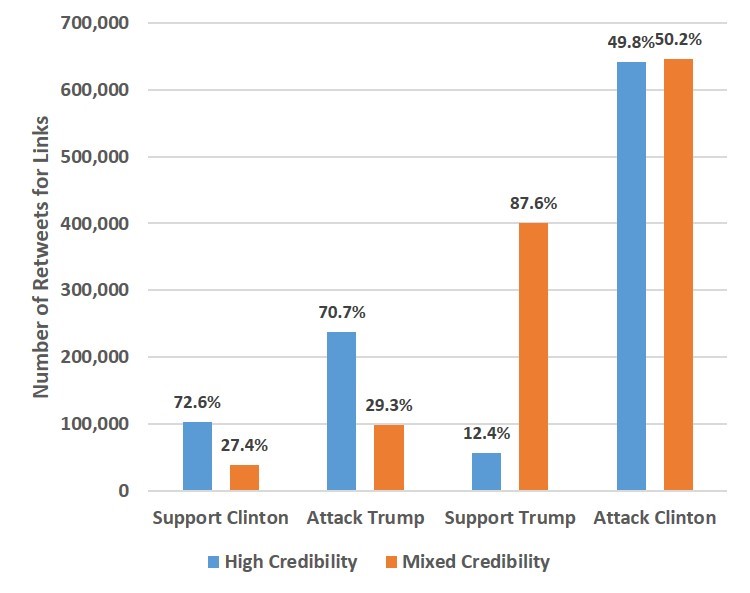}
\label{credibility-graph}
\vspace{-0.5cm}
\caption{Credibility of websites that are linked to from each category}
    \end{minipage}
\vspace{0.2cm}
\end{figure}

Table \ref{table:topDaysForClasses} lists the three days for every class when each class had the largest percentage of tweet volume along with the leading topic. As can be seen, the ``support Clinton'' class not only had the lowest average overall, but it also never exceed 33\% on any given day. All other classes had days when their volume exceeded 60\% of the total retweet volume, with the ``attack Clinton'' class reaching nearly 73\%. The relative volume of ``support Clinton'' retweets peaked after public appearances (promotional video, debate, and TV interview). The ``attack Clinton'' class peaked when her health came into question and after WikiLeaks leaks questioned her integrity. For the ``support Trump'' class, it peaked after polls showed he was ahead and after he announced the building of a wall with Mexico. The ``attack Trump'' class peaked due positions on immigration and women. The most frequent hashtags and terms reflect similar trends.\\ 
\textbf{\textit{Most Frequent Hashtags}} In order to understand the most discussed topics in the viral tweets, Table \ref{table:topHashTagsWithCat} shows the top hashtags that are used in attacking or supporting either candidate divided into categories along with the volume in each category. The top used hashtags for the different labels reveal some stark contrasts between the two candidates. These include: \\
(\emph{\textbf{i}}) The top most frequently appearing hashtags favoring Clinton were those praising her debate performances and attacking Trump. On the Trump side, the most frequently appearing hashtags were those iterating his campaign slogans, encouraging people to vote, and spreading campaign news. \\
(\emph{\textbf{i}})	Hashtags of Trump's campaign slogans appeared nearly 200 times more than Clinton's campaign slogan (\#IAmWithHer). In fact, Trump's campaign slogans category has the most frequently appearing hashtags.\\
(\emph{\textbf{ii}})	Hashtags indicating ``Get out of the vote'' were 4 times more voluminous for Trump than Clinton. Trump and his campaign were more effective in promoting their activities (ex. \#ICYMI -- in case you missed it, \#TrumpRally). \\
(\emph{\textbf{iii}})	Hashtags pertaining to the presidential debate appeared for every class. The support to attack hashtag volume ratio was roughly 3-to-2 for Clinton and 1-to-9 for Trump. Further, the debates were the number one category for the ``support Clinton'' and the ``attack Trump'' classes. It seems that most users thought that she did better in the debates than him. \\
(\emph{\textbf{iv}})	Attacks against Clinton focus on her character and on the WikiLeaks leaks, which cover Clinton's relationship with Wall Street (ex. \#GoldmanSachs), alleged impropriety in the Clinton Foundation (ex. \#FollowTheMoney, \#PayToPlay), mishandling of the Ben Ghazi attack in Libya, Democratic Party primary race against senator Sanders (ex. \#BasementDewellers), the FBI investigation of Clinton (\#AnthonyWeiner), and accusation of witchcraft (\#SpiritCooking). \\
(\emph{\textbf{v}})	Attacks against Trump were dominated by his debate performance. The frequency of hashtags for the next category pertaining to accusations of sexual misconduct (ex. \#TrumpTapes) is one order of magnitude lower than the frequency of those about the debate. This suggests that accusations of sexual misconduct against Trump were not the primary of focus of users. \\
(\emph{\textbf{vi}})	Policy issues such as health care (ex. \#ObamaCare, \#LatinaEqualPay) were eclipsed by issues pertaining to the personalities of the candidates and insults (ex. \#CrookedHillary, \#TangerineNightmare). \\
\textbf{\textit{Most Frequent Terms}} We look at the most frequent terms in the tweets to better understand the most popular topics being discussed. Figure \ref{tag-cloud} shows tag-clouds of the most frequent terms for the different classes, which exhibit similar trends to those in the most frequent hashtags. For example, top terms in the ``attack Clinton'' class include ``crooked'', ``emails'', ``FBI'', ``WikiLeaks'', and ``\#DrainTheSwamp''. Similarly, ``\#DebateNight'' was prominent in the ``support Clinton'' and ``attack Trump'' classes. One interesting word in the ``support Trump'' class is the word ``thank'', which typically appears in Trump authored tweets in conjunction with polls showing Trump ahead (ex. ``Great poll out of Nevada- thank you!''), after rallies (ex. ``Great evening in Canton Ohio-thank you!''), or in response to endorsements (ex. ``Thank you Rep. @MarshaBlackburn!''). Words of thanks appeared in 167 ``support Trump'' tweets that were retweeted more than 1.5 million times compared to 15 ``support Clinton'' tweets that were retweeted 75 thousand times only. Another set of words that do not show in the hashtags are ``pregnancy'' and ``inconvenient'' that come from two tweets that mention that ``Trump said pregnancy is very inconvenient for businesses''. One of these tweets is the most retweeted in the ``attack Trump'' class. \\
\textbf{\textit{Mentions of States}} One of the top terms that appeared in tweets supporting Trump is ``Florida'', Figure \ref{tag-cloud}(c). This motivated us to analyze mentions of states in the tweets of each class. The frequency of state mentions may indicate states of interest to candidates and Twitter users. Figure \ref{state-mentions} lists the number of times each of the 15 most frequently mentioned states. To obtain the counts, we tagged all tweets using a named entity recognizer that is tuned for tweets \cite{ritter2011named}. We automatically filtered entities to obtain geolocations, and then we manually filtered locations to retain state names and city and town names within states. Then, we mapped city and town names to states (ex. ``Grand Rapids'' $\rightarrow$ ``Iowa''). As expected, so-called ``swing states'', which are states that could vote Republican or Democrat in any given election\footnote{\url{https://projects.fivethirtyeight.com/2016-election-forecast/}}, dominated the list. The only non-swing states on the list are New York, Washington, and Texas. Interestingly, the number of mentions in ``support Trump'' tweets far surpasses the counts for all other classes. Most of the mentions of swing states were in tweets authored by Trump indicating Trump rallies being held in these states. This suggests that the Trump campaign effectively highlighted their efforts in these states, and there was significant interest from Twitter users as indicated by the number of retweets. Of the swing states in the figure, Trump won the first six, namely Florida, Ohio, North Carolina, Pennsylvania, Michigan, and Arizona, along with Iowa and Wisconsin. \\
\textbf{\textit{Most Shared Links}} An important debate that surfaced after the US elections was whether fake news affected the election results or not \cite{kucharski2016post,giglietto2016fakes}. Thus, we analyze the credibility of the websites that were most linked to in the tweets for each class. Our analysis shows that 29\% of the viral tweets (21\% of the full retweet volume) contained external links. Table \ref{top-shared-links} shows the top 10 web domains that were linked to in the tweets dataset along with ideological leaning and credibility rating. 
Aside from the official websites of the campaigns (ex. \url{HillaryClinton.com}, \url{Democrats.org}, and \url{DonaldJTrump.com}), the remaining links were to news websites (ex. CNN and NYTimes) and social media sites (ex. Facebook, Instagram, and YouTube). As Table \ref{top-shared-links} shows, the candidates' official websites received the most links. However, it is interesting to note the large difference in focus between both. Clinton's website attacked Trump than it supported Clinton, while Trump's website did the exact opposite. This again shows the differences in strategies and trends between both candidates and their supporters. We checked the leaning and credibility of the news websites on \url{mediabiasfactcheck.com}, a fact checking website which rates news sites anywhere between ``Extreme Left'' to ``Extreme Right'' and their credibility between ``Very Low'' and ``Very High'' with ``Mixed'' being the middle point. We assumed social media sites to have no ideological leaning with a credibility of ``Mixed''. We opted for assigning credibility to websites as opposed to individual stories, because investigating the truthfulness of individual stories is rather tricky and is beyond the scope of this work. Though some stories are easily debunked, such as the Breitbart story claiming that ``Hillary gave an award to a terrorist's wife\footnote{\url{http://bit.ly/2d99lWD/}    \hspace{60pt} $ ^{6}$ \url{http://bit.ly/2tEUxVw}}'', other mix some truth with opinion, stretched truth, and potential lies, such as the Breitbart story that the ``FBI is seething at the botched investigation of Clinton''.
\begin{table}[t]
\tiny
\begin{tabular}{l|r|r|c|c || l|r|r|c|c } 
  \hline
  \multicolumn{5}{c}{Support Clinton} & \multicolumn{5}{c}{Attack Trump} \\ \hline
  Link  & Count & Volume  & Leaning & Credibility & Link  & Count & Volume  & Leaning & Credibility \\ \hline
Hillaryclinton.com  & 63 & 363,153  & Left & --  & Hillaryclinton.com  & 65  & 236,126  & Left  & --\\
Democrats.org  &  8 & 120,026  & Left & -- & WashingtonPost.com & 23 & 102,179  & -2 & High \\
IWillVote.com  & 26 & 101,996  & Left & -- & IWillVote.com  & 11 & 95,126  & Left & --\\
SnappyTV.com  & 14 & 58,926  & N/A & Mixed & SnappyTV.com  & 14 & 86,100  & N/A & Mixed\\
CNN.com  &  6 & 43,355  & -3 & High & Democrats.org & 3 & 67,843  & Left & --\\
WashingtonPost.com  & 10 & 27,264  & -2 & High & Newsweek.com & 11 & 55,663  & -3 & High \\
Medium.com  &  8 & 27,187  & -2 & Mixed & NYTimes.com & 9 & 37,993  & -2 & High\\
BusinessInsider.com  &  4 & 16,010  & -2 & High & CNN.com & 6 & 27,275  & -3 & High\\
NYTimes.com  &  6 & 15,886  & -2 & High & Vox.com  & 4 & 13,992  & -4 & High\\\
YouTube.com  &  2 & 11,564  & N/A & Mixed & Facebook.com  & 2 & 11,973  & N/A & Mixed\\ \hline \hline  

  \multicolumn{5}{c}{Support Trump} & \multicolumn{5}{c}{Attack Clinton} \\ \hline

DonaldJTrump.com  & 77 & 503,375  & right  &   & WikiLeaks.org  & 47 & 406,607  & 2 & High \\
Facebook.com  & 25 & 164,995  & N/A & Mixed  & DailyCaller.com  & 22 & 195,695  & 4 & Mixed \\
WashingtonPost.com  &  7 & 56,855  & -2 & Hight  & FoxNews.com  & 28 & 146,571  & 4 & Mixed \\
Lifezette.com  &  3 & 49,132  & 4 & Mixed  & YouTube.com  & 29 & 107,915  & N/A & Mixed\\
SnappyTV.com  &  4 & 36,663  & N/A & Mixed  & Breitbart.com  & 17 & 102,949  & 5 & Mixed \\
Instagram.com  &  5 & 36,132  & N/A & Mixed  & Politico.com & 11 & 93,165  & -2 & High\\
DailyCaller.com  &  3 & 35,474  & 4 & Mixed  & NYPost.com  & 14 & 92,147  & 3 & Mixed \\
NYPost.com  &  4 & 34,930  & 3 & Mixed  & CNN.com  & 11 & 78,797  & Right & \\
Periscope.tv  & 3 & 22,596  & N/A & Mixed  & Vox.com  & 16 & 77,340  & 3 & High \\
YouTube.com  & 5 & 20,046  & N/A & Mixed & Facebook.com  & 9  & 63,896  & 3 & High \ \\ \hline \hline
  \hline
\end{tabular}
\caption{Top shared links per class -- support/attack Clinton/Trump -- with ideological leaning (-5 extreme left to 5 extreme right) and credibility (high or mixed) according to \url{https://mediabiasfactcheck.com/}}
\label{top-shared-links}
\end{table}
\vspace{-0cm}

Figure \ref{credibility-graph} aggregates all the number of retweets for each category for the websites with credibility of ``High'' and ``Mixed'' -- none of the sites had credibility of ``Very High'', ``Low'', or ``Very Low''. The aggregate number of links for all categories to ``High'' and ``Mixed'' credibility websites was 1.04 million and 1.18 million respectively. The figure shows that a significantly higher proportion of highly credible websites were linked to for the ``support Clinton'' and ``attack Trump'' categories compared to mixed credibility sites. The opposite was true for the ``support Trump'' category, where the majority of links used to support Trump was from mixed creditability websites. This may indicate that Trump supporters were more susceptible to share less credible sources compared to Clinton supporters. High and mixed credibility sites were evenly matched for the ``attack Clinton'' category. WikiLeaks was the foremost shared site for attacking Clinton, which has high credibility. This again highlights the role of WikiLeaks in steering public opinion against Clinton. It worth mentioning that ``The Podesta Emails\footnote{\url{https://wikileaks.org/podesta-emails/}}''  was the most popular link. Thus, though lower credibility links were used to attack Clinton, high credibility links featured prominently. For ideological orientation, left leaning sources featured prominently in the ``support Clinton'' and ``attack Trump'' classes, and right leaning sources featured prominently for the two other classes. The only left leaning sources appearing on the ``support Trump'' and ``attack Clinton'' lists were Washington Post and Politico.\\
\textbf{\textit{Most Retweeted Tweets}} Table \ref{table:top5Retweeted} lists the top 5 most retweeted\footnote{The number of retweets in the table are taken in November 2016} tweets for each class. They illustrate the trends to those exhibited in our previous analysis, namely: ``support Clinton'' tweets are led by talk about debates and attacks against Trump; the most retweeted ``attack Clinton'' tweets are from WikiLeaks; Trump campaign slogans appear atop of ``support Trump'' tweets; and the most retweeted attack Trump'' tweets cover demeaning statements against women and minorities and mockery of Trump. 
An interesting observation is that the most retweeted tweet attacking Trump came from a Nigerian account with screen name ``Ozzyonce'' who had less than 2,000 followers at the time when she posted the tweet. Nonetheless, this tweet received over 150K retweets in one day. \\
\textbf{\textit{Most Retweeted Accounts}}
Table \ref{table:topRetweeted} lists the most retweeted accounts for each class. As expected, the most retweeted accounts for ``support'' and ``attack'' classes were those of the candidates themselves and their rivals respectively. Notably:\\ 
(\emph{\textbf{i}}) Clinton authored 10\% more ``attack Trump'' tweets (with 35\% more volume) than ``support Clinton''. In contrast, Trump authored 81\% more ``support Trump'' tweets (with 38\% more volume) than ``attack Clinton'' tweets. This suggests that Clinton expended more energy attacking her opponent than promoting herself, while Trump did the exact opposite. Trump campaign staffers and accounts, such as Kellyanne Conway, Dan Scavino Jr., and Official Team Trump, were slightly more active in the ``attack Clinton'' class than the ``support Trump'' class.\\
(\emph{\textbf{ii}}) Trump Campaign accounts featured prominently in ``support Trump'' and ``attack Clinton'' classes, capturing 7 out of 10 and 4 out of 10 top spots for both classes respectively. In contrast, Clinton campaign account captured 2 out of 10 and 1 out of 10 top spots for ``support Clinton'' and ``attack Trump'' classes respectively. Top Clinton aides like Huma Abedin and John Podesta were absent from the top list, suggesting a more concerted Trump campaign Twitter strategy.\\
(\emph{\textbf{iii}}) Though Clinton authored 48\% more tweets attacking her rival than tweets Trump authored attacking her, his attack tweets led to 34\% more retweet volume. This suggests that his attacks were more effective than her attacks. Trump tweets were more retweeted on average than Clinton's tweets with an average of 11,195 and 6,120 retweets per tweet for both respectively. Highlighting WikiLeaks' role in the election, WikiLeaks was the second most retweeted account attacking Clinton and had four times as much retweet volume than the next account.

\begin{table}[t]
\tiny
\begin{center}
\begin{tabular}{r|c|l|r}
\hline
Date & Author & Text & Count \\ \hline \hline
\multicolumn{4}{c}{support Clinton} \\ \hline
9/27	&	Jerry Springer	& \begin{minipage}[t]{.7\columnwidth}	Hillary Clinton belongs in the White House. Donald Trump belongs on my show.	\end{minipage}	&	78,872	\\
9/27	&	Hillary Clinton	& \begin{minipage}[t]{.7\columnwidth}	RT this if you re proud to be standing with Hillary tonight. \#debatenight https://t.co/91tBmKxVMs	\end{minipage}	&	72,443	\\
10/10	&	Erin Ruberry	& \begin{minipage}[t]{.7\columnwidth}	Hillary is proof a woman can work hard rise to the top of her field still have to compete against a less qualified man for the same job.	\end{minipage}	&	72,167	\\
10/20	&	Hillary Clinton	& \begin{minipage}[t]{.7\columnwidth}	RT if you're proud of Hillary tonight. \#DebateNight \#SheWon https://t.co/H7CJep7APX	\end{minipage}	&	72,150	\\
10/8	&	Richard Hine	& \begin{minipage}[t]{.7\columnwidth}	Trump: Don t judge me on the man I was 10 years ago. But please judge Hillary on the man her husband was 20 years ago \#TrumpTapes	\end{minipage}	&	66,817	\\
\hline \hline
\multicolumn{4}{c}{attack Clinton} \\ \hline
11/1	&	WikiLeaks	& \begin{minipage}[t]{.7\columnwidth}	No link between Trump Russia No link between Assange Russia But Podesta Clinton involved in selling 20\% of US uranium to Russia	\end{minipage}	&	51,311	\\
10/14	&	WikiLeaks	& \begin{minipage}[t]{.7\columnwidth}	Democrats prepared fake Trump grope under the meeting table Craigslist employment advertisement in May 2016 https://t.co/JM9JMeLYet	\end{minipage}	&	45,348	\\
10/3	&	WikiLeaks	& \begin{minipage}[t]{.7\columnwidth}	Hillary Clinton on Assange Can t we just drone this guy -- report https://t.co/S7tPrl2QCZ https://t.co/qy2EQBa48y	\end{minipage}	&	45,233	\\
11/4	&	Donald J. Trump	& \begin{minipage}[t]{.7\columnwidth}	If Obama worked as hard on straightening out our country as he has trying to protect and elect Hillary we would all be much better off!	\end{minipage}	&	42,331	\\
11/8	&	Cloyd Rivers	& \begin{minipage}[t]{.7\columnwidth}	To everyone who wants to vote for Hillary just to keep Trump from becomin President watch this... https://t.co/TuUpD7Qgcg	\end{minipage}	&	41,047	\\
\hline \hline
\multicolumn{4}{c}{support Trump} \\ \hline
11/8	&	Donald J. Trump	& \begin{minipage}[t]{.7\columnwidth}	TODAY WE MAKE AMERICA GREAT AGAIN!	\end{minipage}	&	352,140	\\
11/5	&	Donald J. Trump	& \begin{minipage}[t]{.7\columnwidth}	MAKE AMERICA GREAT AGAIN!	\end{minipage}	&	60,718	\\
10/8	&	Donald J. Trump	& \begin{minipage}[t]{.7\columnwidth}	Here is my statement. https://t.co/WAZiGoQqMQ	\end{minipage}	&	52,887	\\
11/7	&	Immigrants4Trump	& \begin{minipage}[t]{.7\columnwidth}	If you make this go viral Trump will win. It s about 2 minutes that makes the choice in this election crystal clear https://t.co/CqMY4CSJbp	\end{minipage}	&	43,627	\\
10/10	&	Mike Pence	& \begin{minipage}[t]{.7\columnwidth}	Congrats to my running mate @realDonaldTrump on a big debate win! Proud to stand with you as we \#MAGA.	\end{minipage}	&	42,178	\\
\hline \hline
\multicolumn{4}{c}{attack Trump} \\ \hline
9/15	&	Ozzyonce	& \begin{minipage}[t]{.7\columnwidth}	Donald Trump said pregnancy is very inconvenient for businesses like his mother s pregnancy hasn't been inconvenient for the whole world.	\end{minipage}	&	152,756	\\
10/26	&	Bailey Disler	& \begin{minipage}[t]{.7\columnwidth}	Good morning everyone especially the person who destroyed Donald Trump s walk of fame star https://t.co/IcBthxMPd9	\end{minipage}	&	124,322	\\
10/21	&	Stephen King	& \begin{minipage}[t]{.7\columnwidth}	My newest horror story: Once upon a time there was a man named Donald Trump and he ran for president. Some people wanted him to win.	\end{minipage}	&	121,635	\\
10/10	&	Kat Combs	& \begin{minipage}[t]{.7\columnwidth}	Trump writing a term paper: Sources Cited: 1. You Know It 2. I know It 3. Everybody Knows It	\end{minipage}	&	105,118	\\
10/8	&	Es un racista	& \begin{minipage}[t]{.7\columnwidth}	Anna for you to sit here call Trump a racist is outrageous Anna: OH?! Well lemme do it again in 2 languages! https://t.co/nq4DO7bN7J	\end{minipage}	&	102,063	\\ \hline
\end{tabular}
\caption{Top retweeted tweets for each class -- support/attack Clinton/Trump.}
\label{table:top5Retweeted}
\vspace{-0.4cm}
\end{center}
\end{table}

\begin{table}[ht]
\tiny
\begin{center}
\setlength\tabcolsep{2.5pt}
\begin{tabular}{l|r|r|l|r|r}
\hline
\multicolumn{3}{c|}{support Clinton} & \multicolumn{3}{c}{attack Trump} \\ \hline
Account	&	Count	&	Volume	&	Account	&	Count	&	Volume	\\ \hline
\textit{Hillary Clinton}		&	 331		&	 2,025,821		&		\textit{Hillary Clinton}		&	363	&		2,698,209	\\
President Obama		&	 4		&	 122,947		&		Bernie Sanders		&	11	&		304,860	\\
\textit{Senator Tim Kaine}		&	15	&		84,245		&		Ozzyonce		&	1	&		152,756	\\
Jerry Springer		&	 1		&	 78,872		&		Bailey Disler		&	1	&		124,322	\\
Erin Ruberry		&	 1		&	 72,167		&		Stephen King 	&	2	&		121,635	\\
Richard Hine		&	1	&	 66,817		&		Kat Combs		&	1	&		105,118	\\
Bernie Sanders		&	7	&	 46,180		&		Es un racista		&	1	&		102,063	\\
CNN		&	6	&	 41,983		&		Rob Fee		&	1	&		99,401	\\
Funny Or Die		&	1	&	 27,909		&		Jerry Springer		&	1	&		78,872	\\
Channel 4 News		&	1	&	 27,409		&		Master of None 	&	 1		&		67,690	\\ \hline
\multicolumn{3}{c|}{Support Trump} & \multicolumn{3}{c}{attack Clinton} \\ \hline				
	\textit{Donald J. Trump}		&	446	&		4,992,845		&		\textit{Donald J. Trump}		&	246	&		3,613,025		\\
	\textit{Kellyanne Conway}		&	51	&			199,511		&		WikiLeaks		&	141	&		1,454,903		\\
	\textit{Mike Pence}		&	36	&			195,824		&		\textit{Kellyanne Conway} 	&	 92		&		349,025		\\
	\textit{Dan Scavino Jr.}		&	40	&			181,601		&		Paul Joseph Watson	 	&	78	&		297,273		\\
	\textit{Official Team Trump}		&	14	&			141,289		&		\textit{Official Team Trump}		&	23	&		150,932		\\
	\textit{Donald Trump Jr.}		&	20	&			112,835		&		\textit{Donald Trump Jr.} 	&	 34		&		126,744		\\
	\textit{Eric Trump}		&	8	&			79,387		&		Jared Wyand		&	19	&		85,742		\\
	Immigrants4Trump		&	4	&			52,256		&	 Cloyd Rivers		&	7	&		84,063		\\
	Cloyd Rivers		&	2	&			45,493		&		Juanita Broaddrick		&	4	&		83,903		\\
	Paul Joseph Watson		&	10	&		38,474		&		James Woods		&	16	&		78,719		\\ \hline
\end{tabular}
\caption{Top retweeted accounts per class -- support/attack Clinton/Trump.  Accounts of candidates and campaign affiliates are italicized.}
\label{table:topRetweeted}
\end{center}
\end{table}

\section{Discussion}
\vspace{-0.2cm}
Our analysis 
contrasts differences in 
support for either candidate, namely: \\ 
\textbf{Popularity:} While Trump received more negative coverage than Clinton in mainstream media \cite{van2017leading}, Trump benefited from many more tweets that are either supporting him or attacking his rival. In fact, 63\% of the volume of viral content on US election were in his favor compared to only 37\% in favor of Clinton. Similarly, on 85\% of the days in the last two months preceding the election, Trump had more tweets favoring him than Clinton. This observation shows the gap between the trends of social media and traditional news media.\\
\textbf{Positive vs negative attention:} The volume of tweets attacking and supporting Trump were evenly matched, while the volume of tweets attacking Clinton outnumbered tweets praising her by a 3-to-1 margin. Given the importance of social media in elections, this may have been particularly damaging.\\
\textbf{Support points:} Trump campaign accounts featured more prominently in the top retweeted accounts supporting their candidate. Support came in the form of promoting his slogans, urging supporters to vote, and featuring positive polls and campaign news. Conversely, Clinton support came mostly in the form of contrasting her to her rival, praise for her debate performance against Trump, and praise for her attacks on Trump. In fact, viral tweets from her campaign account were attacking Trump more than promoting her. Unfortunately for her, she was framed in reference to her rival, and  research has shown that debates have a dwindling effect election outcomes as the election day draws closer\cite{benoit2003meta,erikson2012timeline,hillygus2003voter}. Consequently, her post-debate surges in volume of attacks against Trump eclipsed surges of support for her.\\
\textbf{Attack points:} Both candidates were attacked on different things. Trump was mainly attacked on his debate performance, eclipsing attacks related to his scandals, such as the lewd Access Hollywood tape. Luckily for him, debates have a diminishing effect on voters as election day draws near \cite{benoit2003meta,erikson2012timeline,hillygus2003voter}. Conversely for Clinton, persistent allegations of corruption and wrongdoing triggered by WikiLeaks and the FBI investigation of her emails dominated attacks against her. These attacks may have led to her eventual loss, with polls suggesting that ``email define(d) Clinton''\footnote{\url{gallup.com/poll/185486/email-defines-clinton-immigration-defines-trump.aspx}}.\\
\textbf{Message penetration:} Trump's slogan, ``Make America Great Again'', had a far greater reach than that of his rival, ``Stronger Together''. Similarly, his policy positions and agenda items, such as the proposal to build a wall with Mexico, attracted significantly more attention than those of Clinton, where her proposed policy positions received very little mention. \\
\textbf{Geographical focus:} Trump and his supporters effectively promoted his campaign's efforts in swing state, with frequent mentions of rallies and polls from these states along with messages of thanks for people turning-out for his rallies. The volume of tweets mentioning swing states and supporting Trump were typically two orders of magnitude larger than similar tweets supporting Clinton. This might have contributed to the narrow victory he achieved in many of them.\\
\textbf{Low credibility links:} Trump supporters were more likely to share links from websites of questionable credibility than Clinton supporters. However, WikiLeaks, which has high credibility, was the most prominent source attacking Clinton.

To better understand the presented results, a few limitations that need to be considered.  First, the top 50 viral tweets do not have to be representative of the whole collection. Nonetheless, they still represent over 40\% of the tweets volume on the US elections during the period of the study. Second, the results are based on tweets collected from TweetElect. Although it is highly robust, the site uses automatic filtering methods that are not perfect \cite{magdy2014adaptive}. Therefore, there might be other relevant viral tweets that were not captured by the filtering method. Lastly, measuring support for a candidate using viral tweets does not have to represent actual support on the ground for many reasons. Some of these reasons include the fact that demographics of Twitter users may not match the general public, more popular accounts have a better chance of having their tweets go viral, or either campaign may engage in astroturfing, in which dedicated groups or bots may methodically tweet or retweet pro-candidate messages \cite{bessi2016social}. 

\vspace{-0.5cm}
\section{Conclusion}
\vspace{-0.3cm}
In this paper, we presented quantitative and qualitative analysis of the top retweeted tweets pertaining to the US presidential elections from September 1, 2016 to election day on November 8, 2016. For everyday, we tagged the top 50 most retweeted tweets as supporting/attacking either candidate or as neutral/irrelevant. Then we analyzed the tweets in each class from the perspective of: general trends and statistics; most frequent hashtags, terms, and locations; and most retweeted accounts and tweets. Our analysis highlights some of the differences between the social media strategies of both candidates, the effectiveness of both in pushing their messages, and the potential effect of attacks on both. We show that compared to the Clinton campaign, the Trump campaign seems more effective in: promoting Trump's messages and slogans, attacking and framing Clinton, and promoting campaign activities in ``swing'' states.
For future work, we would like to study the users who retweeted the viral tweets in our study to ascertain such things as political leanings and geolocations. This can help map the political dynamics underlying the support and opposition of both candidates.


%
%
\bibliographystyle{splncs03} 
\bibliography{bibliography,bib}






\end{document}